\documentclass[PRL,twocolumn,showpacs,superscriptaddress,preprintnumbers,nofootinbib,amsmath,amssymb]{revtex4}

\usepackage{graphicx}
\usepackage{dcolumn}
\usepackage{bm}

\usepackage{epstopdf}

\usepackage{amssymb,amsmath,amsthm,mathrsfs}

\begin{document}

\newcommand{\Eq}[1]{Eq. \ref{eqn:#1}}
\newcommand{\Fig}[1]{Fig. \ref{fig:#1}}
\newcommand{\Sec}[1]{Sec. \ref{sec:#1}}

\newcommand{\PHI}{\phi}
\newcommand{\vect}[1]{\mathbf{#1}}
\newcommand{\Del}{\nabla}
\newcommand{\unit}[1]{\mathrm{#1}}
\newcommand{\x}{\vect{x}}
\newcommand{\ScS}{\scriptstyle}
\newcommand{\ScScS}{\scriptscriptstyle}
\newcommand{\xplus}[1]{\vect{x}\!\ScScS{+}\!\ScS\vect{#1}}
\newcommand{\xminus}[1]{\vect{x}\!\ScScS{-}\!\ScS\vect{#1}}
\newcommand{\diff}{\mathrm{d}}

\newcommand{\be}{\begin{equation}}
\newcommand{\ee}{\end{equation}}
\newcommand{\bea}{\begin{eqnarray}}
\newcommand{\eea}{\end{eqnarray}}
\newcommand{\vu}{{\mathbf u}}
\newcommand{\ve}{{\mathbf e}}
\newcommand{\vk}{{\mathbf k}}
\newcommand{\vx}{{\mathbf x}}
\newcommand{\vy}{{\mathbf y}}
\newcommand{\bx}{{\bf x}}
\newcommand{\bk}{{\bf k}}
\newcommand{\br}{{\bf r}}
\newcommand{\bq}{{\bf q}}
\newcommand{\bp}{{\bf p}}
\newcommand{\bv}{{\bf v}}
\newcommand{\TT}{{\rm TT}}
\newcommand{\lqn}[1]{\lefteqn{#1}}
\newcommand{\GW}{{_{\rm GW}}}
\newcommand{\lb}{\left\lbrace}
\newcommand{\rb}{\right\rbrace}
\newcommand{\ab}{{\alpha\beta}}

\newcommand{\uden}{\underset{\widetilde{}}}
\newcommand{\den}{\overset{\widetilde{}}}
\newcommand{\denep}{\underset{\widetilde{}}{\epsilon}}

\newcommand{\dd}{\diff}
\newcommand{\fr}{\frac}
\newcommand{\del}{\partial}
\newcommand{\eps}{\epsilon}
\newcommand\CS{\mathcal{C}}

\def\be{\begin{equation}}
\def\ee{\end{equation}}
\def\ben{\begin{equation*}}
\def\een{\end{equation*}}
\def\bea{\begin{eqnarray}}
\def\eea{\end{eqnarray}}
\def\bal{\begin{align}}
\def\eal{\end{align}}

\def\TT{{\rm TT}}
\def\GW{{_{\rm GW}}}
\newcommand{\mn}{\mu\nu}
\newcommand{\nn}{\nonumber \\}


\title{Imprints of the Standard Model in the Sky:\\ A Gravitational Wave Background from the decay of the Higgs after inflation}

\author{Daniel G. Figueroa\vspace*{.1mm}\\{\it D\'epartement de Physique Th\'eorique and Center for Astroparticle Physics,\vspace*{-.5mm}\\ Universit\'e de Gen\`eve, 24 quai Ernest Ansermet, CH--1211 Gen\`eve 4, Switzerland}}


\begin{abstract}
The existence of the Standard Model (SM) Higgs implies that a gravitational wave (GW) background is generated by the decay products of the Higgs, soon after the end of inflation. 
Theoretically, all Yukawa and $SU(2)_{\rm L}$ gauge couplings of the SM are imprinted as features in the GW spectrum. In practice, the signal from the most strongly coupled species dominates, rendering inaccesible the information on the other species. If detected, this background could be used for measuring properties of high-energy particle physics, including beyond the SM. To achieve this goal, new high frequency GW detection technology is required, beyond that of currently planned detectors.
\end{abstract}

\keywords{cosmology} \pacs{To be done}

\maketitle


\begin{center}
{\bf I. Introduction}\vspace*{-2mm}
\end{center}
Compelling evidence~\cite{Planck2013Overview} strongly supports the idea of inflation, a phase of accelerated expansion in the eary Universe, driven by some vaccuum-like energy density. After inflation, reheating follows, converting all the energy available into different particle species, which eventually thermalize and dictate the expansion of the Universe. The specific realisation and energy scale of inflation are however uncertain. The details of reheating and of the processes during the first stages of the thermal era are also not known. In general, these are expected to be high energy phenomena which cannot be probed by particle accelerators like the Large Hadron Collider. Only cosmic relics produced during this epoch can be used to probe the physics of these primeval instants. 

One of the most promising relic are gravitational waves (GWs). Once produced, GWs decouple and propagate at the speed of light, carrying information about the source that originated them. A background of GWs is expected from inflation which, if detected, will reveal the energy scale of inflation~\cite{StarobinskyGW1979}. Several backgrounds of GWs are also expected from the post-inflationary early Universe period, like preheating~\cite{GWs1997KhlebnikovTkachev,GWs2006EastherLim,GWsPRL2007FigueroaGarciaBellido,
GWsPRD2007FigueroaGarciaBellidoSastre,GWs2007DufauxEtAl,GWs2009HybridInfDufauxEtAl}, phase transitions~\cite{Kamionkowski:1993fg,Caprini:2007xq,Huber:2008hg,Hindmarsh:2013xza}, or cosmic defects~\cite{Vachaspati:1984gt,GWsGeneral2004DamourVilenkin,GWsSelfOrderingScalarFieldsFenuEtAl2009,
GWsAbelianHiggsHybridPreheatingDufauxEtAl2010,GWsUniversalBackgroundFromCDsFigueroaEtAl2012}. Each of these backgrounds has a characteristic spectrum depending on the high energy process that generated them. If detected, GWs will provide direct information about the physics of that epoch. 

In this letter we describe a GW background generated after the end of inflation, created due to the decay of the Standard Model (SM) Higgs. ATLAS and CMS have firmly stablished~\cite{ATLAS:2012ae,CMS:2012tx} the existence of the Higgs, with a mass of 125-126 GeV. 
We ignore however the role of the Higgs in the early Universe or, more precisely, during inflation. Generically, one expects that the Higgs played no dynamical role during inflation, though in principle it could also be responsible for it if a non-minimal coupling to gravity is present~\cite{Bezrukov:2007ep}. The two situations share in common that at the end of inflation, the Higgs is a condensate which oscillates around the minimum of its effective potential. This gives rise to particle creation through non-perturbative parametric effects~\cite{Traschen:1990sw,Kofman:1994rk,Kofman:1997yn,Greene:1998nh,Giudice:1999fb,Greene:2000ew,GarciaBellido:2000dc,Peloso:2000hy}. All particle species coupled directly to the Higgs are then created out-of-equilibrium. The transverse-traceless (TT) part of the energy-momentum tensor of the Higgs decay products represents a source of GWs. As a result, each of the produced species contributes to generate a background of GWs.  

In this paper we compute the spectral shape of this background. We find that every field coupled to the Higgs leaves an imprint in the GW spectrum, but in practice the signal from the most strongly coupled species dominates. 
We discuss the implications of this result as a probe of particle couplings in high-energy physics. We focus on the situation when the Higgs plays no dynamical role during inflation. We consider also, albeit more briefly, the case when the Higgs is responsible for inflation. All through the text $a(t)$ is the scale factor, $t$ conformal time, $\hbar = c = 1$, and $M_p$ = $1/8\pi G \simeq 2.44\times10^{18}$ GeV is the reduced Planck mass, with $G$ the gravitational constant.\\
 
\begin{center}
{\bf II. Higgs oscillations after inflation}
\end{center}
Let us characterize inflation as a {\it de Sitter} period with Hubble rate $H_{\rm e} \gg v \equiv 246$ GeV. In the unitary gauge $\Phi = \varphi/\sqrt{2}$,  the large field effective potential of the SM Higgs is $V = \lambda(\mu)\varphi^4/4$, with $\lambda$ the running self-coupling at the renormalization scale $\mu = \varphi$~\cite{Espinosa:2007qp}. If the Higgs is decoupled (or weakly coupled) from (to) the inflationary sector, it plays no dynamical role during inflation, behaving as a light spectator field independently of its initial amplitude~\cite{DeSimone:2012qr,Enqvist:2013kaa}. The Higgs then performs a random walk at superhorizon scales, reaching quickly an equilibrium distribution $P_{\rm eq} \propto \exp\lbrace{-(2\pi^2\lambda/3)(\varphi/H_{\rm e})^4}\rbrace$~\cite{Starobinsky:1994bd}, with variance $\langle \varphi^2 \rangle \simeq 0.13\lambda^{-1/2}H_{\rm e}^2$. A typical Higgs amplitude at the end of inflation is $\varphi_{\rm e} \sim \mathcal{O}(0.1)H_{\rm e}/\lambda_{e}^{1/4}$, with $\lambda_e = \lambda(\varphi_{\rm e})$. More concretely, $\varphi_{\rm e}$ ranges between $0.01H_{\rm e}/\lambda_{e}^{1/4}$ and $H_{\rm e}/\lambda_{e}^{1/4}$ with $\sim 98\%$ probability.

The running of the Higgs self-coupling shows that $\lambda(\mu_c) = 0$ at at some scale $\mu_c$, above which $\lambda(\mu)$ becomes negative~\cite{Espinosa:2007qp,Degrassi:2012ry}. For the best fit SM parameters $\mu_c \sim 10^{11}$ GeV, though it can be pushed up to $M_p$ considering the top quark mass $3\sigma$ below its best fit. To guarantee the stability of the SM all the way up to inflation, we demand $\lambda_{\rm e} > 0$. This, together with the uncertainty of $H_e$, allows us to effectively consider $\lambda_e$ as a free parameter, simply constrained by $0 < \lambda_e \ll 1$.

The Higgs slowly starts rolling down its potential as soon as inflation ends. Depending on the inflationary sector (here unspecified), the universe just after inflation can be matter-dominated (MD), radiation-dominated (RD), or in-between. The Hubble rate $H$ decreases in any case faster than $\varphi$, eventually becoming $H^2 < V''$. From then on, the Higgs starts oscillating around $\varphi = 0$, with an initial amplitude $\varphi_{\rm I} = H_{\rm I}/\lambda_{\rm I}^{1/2} (< \varphi_{\rm e})$, where $\lambda_{\rm I}$ $\equiv$ $\lambda(E_{\rm I})$ $\gtrsim$ $\lambda_{\rm e}$, $E_{\rm I} \sim (M_pH_{\rm I})^{1/2}$. The initial velocity can be read from the slow-roll condition, $d\varphi_{\rm I}/dt = - V'/2H_{\rm I}$. Rescaling the Higgs amplitude as $h \equiv a\varphi/\varphi_i$, and defining the time variable $d\tau \equiv \sqrt{\lambda_{\rm I}}\varphi_i dt$ (so $\dot{} \equiv d/d\tau$ from now on), the initial conditions read $h_i =1, \dot h_i = -1/2$. 
The Higgs condensate oscillates then according to
\begin{equation}\label{eq:HiggsEOMspectator}
\ddot h(\tau) + h^3(\tau) = ({\ddot a/ a})h(\tau)\,.
\end{equation} 
\vspace{0.02mm}

\begin{center}
{\bf III. GWs from the Higgs decay products}
\end{center}
The oscillations of the Higgs condensate have a striking consequence: everytime $\varphi$ passes through zero, all particle species coupled to the Higgs are created out-of-equilibrium through non-perturbative effects~\cite{Traschen:1990sw,Kofman:1994rk,Kofman:1997yn,Greene:1998nh,Giudice:1999fb,Greene:2000ew,GarciaBellido:2000dc,Peloso:2000hy}. This occurs much faster than particle production from the perturbative decay of the Higgs~\cite{Enqvist:2013kaa}. In particular, $SU(2)_{\rm L}$ gauge bosons and all charged fermions of the SM are created at the first and sucessive Higgs zero crossings. The energy momentum tensor $T_{\mu\nu}$ of the excited species represents an anisotropic stress over the background and, consequently, its TT part acts as a source of GWs. Thus, all species excited due to Higgs oscillations, are expected to generate GWs. In this letter we focus on the GW production from the SM charged fermions. Nonetheless we note that gauge bosons are also expected to produce GWs, see Section V. 

Let us define now, for later convenience, the times $t_{\rm e}$, $t_{\rm I}$, $t_{\rm F}$ and $t_{\rm RD}$, as the end of inflation, the start of the Higgs oscillations, the end of GW production, and the first moment when the Universe becomes RD. 

It was shown recently that parametrically excited fermions can generate very efficiently GWs~\cite{Enqvist:2012im,Figueroa:2013vif}. Let $(y/\sqrt{2})\varphi\bar\psi\psi$ be the Yukawa interaction of a given fermion species $\psi$ with the Higgs, with $y$ the Yukawa coupling strength. We can decompose the fermionic field as $\psi(\bx,t) = (2\pi)^{-3}\int d\bk e^{-i\bk\bx}\lbrace  a_{\bk,r}{\tt u}_{\bk,r}(t) + b^\dag_{-\bk,r}{\tt v}_{\bk,r}(t) \rbrace$, with ${r = 1,2}$ polarization indices, ${\tt v}_{\bk,r} \equiv i\gamma_0\gamma_2\bar{{\tt u}}_{\bk,r}$, ${\tt u}_{\bk,r} = (u_{\bk,+}S_r~,~u_{\bk,-}S_r)^{\rm T}$, $S_{1}, S_{2}$ eigenvectors of the helicity operator, and $a_{r}, b_{r}$ standard creation/annihilation operators obeying the usual anti-commutation relations. 
The fermion mode functions then obey the equations
\begin{equation}\label{eq:FermionsEOM}
\ddot u_{k,\pm} + \left(\kappa^2 + q h^2 \pm i\sqrt{q}\dot h\right)u_{k,\pm} = 0\,,~~~~q \equiv {y^2/\lambda_{\rm I}}\,,
\end{equation}
where $\kappa \equiv k/H_{\rm I}$, $q$ is a 'resonance' parameter, and $u_{k,\pm}(t_{\rm I}) \equiv [1\pm 1/(1+\kappa^2/q)^{1/2}]^{1/2}$ and ${\dot u}_{k,\pm}(t_{\rm I}) \equiv i[\kappa u_{k,\mp}(t_{\rm I}) \mp q^{1/2}u_{k,\pm}(t_{\rm I})]$ guarantee an initially vanishing fermion number density. Solving Eq.~(\ref{eq:HiggsEOMspectator}) we find $h(t)$, plug it into Eq.~(\ref{eq:FermionsEOM}), and then solve for the mode functions $u_{k,\pm}(t)$. This scheme is consistent as long as the backreaction from fermions into the Higgs is not relevant. 

The normalized energy density spectrum of GWs, $\Omega_\GW(k,t) \equiv {1\over\rho_c}\frac{d\rho_{\GW}}{d\log k}$, $\rho_c = {3H^2\over 8\pi G}$ the critical energy density, generated by a fermionic field with mode functions $u_{k,\pm}(t)$, is given by~\cite{Figueroa:2013vif}
\begin{eqnarray}
\begin{array}{c}
\Omega_\GW(k,t) = {4\over 3\pi^3}\frac{G^2k^{3}}{H^2a^{4}(t)}\int \hspace*{-1mm} d\vec p\,p^{2}\sin^{2}\hspace*{-0.6mm}\theta\,\big(\big|I_{(c)}\big|^{2} + \big|I_{(s)}\big|^{2}\big)\vspace*{+2mm}\,,\\
I_{(c)}(\vec k,\vec p,t) \equiv \hspace*{-.5mm}\int_{t}\frac{dt'}{a(t')}\cos(kt')\mathcal{K}_{\rm reg}(k,p)W_{k,p}(t')\,,
\end{array}\label{eq:GW_spectra_Reg}
\end{eqnarray}
with $W_{k,p} \equiv (u_{k-p,+}u_{p,+}-u_{k-p,-}u_{p,-})$, $\mathcal{K}_{\rm (reg)}(k,p) \equiv 2(n_{k-p}n_{p})^{1/2}$, $n_p, n_{k-p}$ the occupation numbers, and $I_{(s)}$ analogously defined as $I_{(c)}$ but with $\sin(kt)$. Note that parametric creation of fermions excites modes up to a given cut-off scale $k_* \simeq q^{1/4}H_{\rm I}$, i.e.~only infrared (IR) modes ($k \lesssim k_*$) are excited, whilst ultraviolet (UV) modes ($k \gtrsim k_*$) remain in vaccuum. The contribution from the UV modes diverges and must be subtracted ('regularized'). The kernel $\mathcal{K}_{\rm reg}(p,k)$ appears precisely due to the regularization of the anisotropic-stress~\cite{Figueroa:2013vif}, acting as a IR filter which suppresses the UV contribution, i.e.~$\mathcal{K}_{\rm reg}(p,k) \rightarrow 0$ when $p, k \gg k_*$. 

\begin{figure}[t]
\begin{center}
\includegraphics[width=6cm]{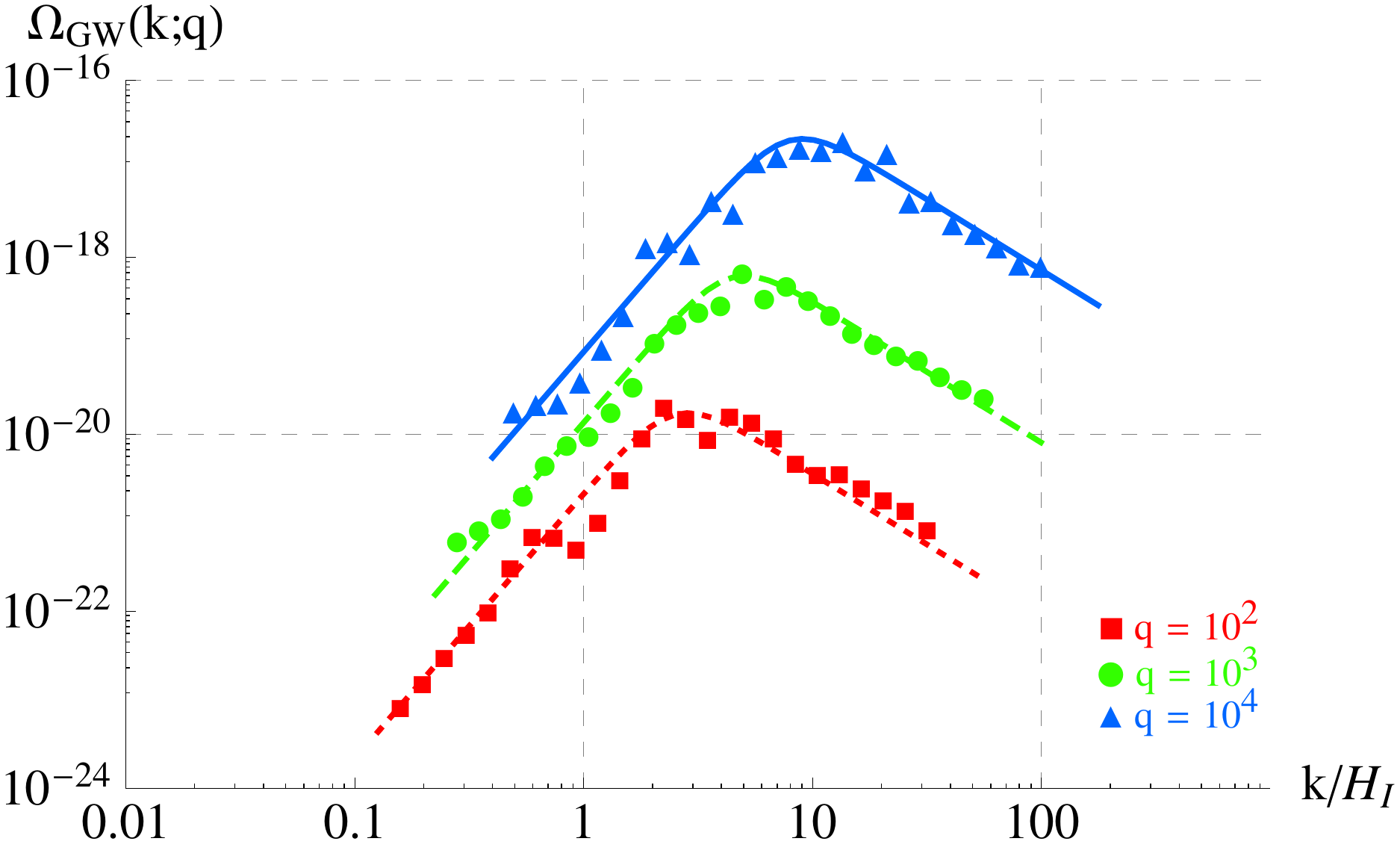}\vspace*{5mm}
\includegraphics[width=6cm]{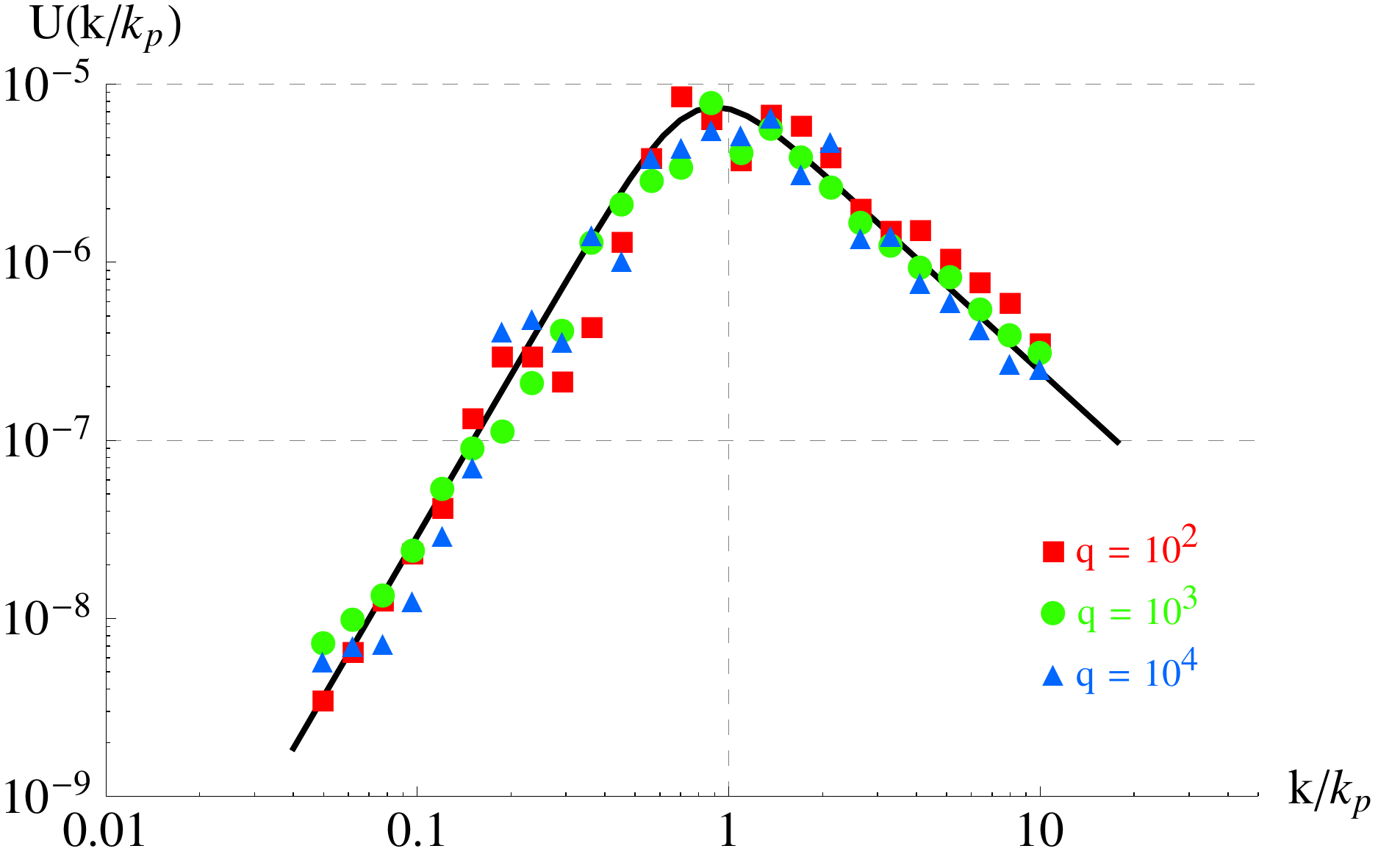}
\end{center}
\caption{Top: Three GW spectra calculated for the resonance parameters $q = 10^2, 10^3, 10^4$ in RD and for $H_{\rm I} = H_{\rm I}^{<}$. Bottom: Universal function $\mathcal{U}(k/k_p)$, obtained from the GW spectra calculated above. As expected, the spectra peak at $k_p \sim q^{1/4}H_{\rm I}$, signaled by the maximum of $\mathcal{U}(x)$ at $x \sim 1$. Similar plots are obtained for the MD case.
}\label{fig:FigureI}
\end{figure}

Since the fermionic spectrum has a hard cut-off at $k_*$, the GW spectrum 
must be peaked at a scale $k_p \sim k_*$, with a $k^3$ slope for $k \ll k_*$, and a decaying UV tail at $k \gg k_*$ (due to the fermion occupation number suppresion). In Figure~\ref{fig:FigureI} several GW spectra are shown as an example, computed for $q = 10^2, 10^3, 10^4$ in RD. All spectra show the expected behavior, the $k^3$ IR tail, a peak at $k_p \sim q^{1/4}H_{\rm I}$, and a decaying amplitude at $k \gg k_p$. The UV tails fit well with a power-law $\propto k^{-1.5}$, but this should be taken with care given the limited momenta range probed. The amplitude of the GW peak is expected to scale as $\Omega_\GW^{\rm (p)}$ $\equiv$ $\Omega_\GW(k=k_p) \propto$ $q^{(3+\delta)/2}$~\cite{Figueroa:2013vif}, with $\delta < 1$ a small correction depending on the fermion number suppression details at $k > k_*$. Numerically we find $\Omega_\GW^{\rm (p)} \propto q^{1.55}$ for both RD/MD, so $\delta \simeq 0.1$. Denoting as $w$ the effective equation of state parameter characterizing the expansion history betwen $t_{\rm I}$ and $t_{\rm RD}$, the GW spectrum for a given resonance parameter $q \ge 1$, can be parametrized as
\begin{eqnarray}
\begin{array}{c}
\Omega_\GW(k,t_{\rm F};q) = q^{1.55}\,\mathcal{U}(k/k_p) \times ({H_{\rm I}/M_p})^{{4}}\,({a_{\rm I}/a_{\rm F}})^{{1-3w}}\vspace*{2mm}\\
\mathcal{U}(x) \equiv ~\mathcal{U}_1\cdot x^3/(\alpha + \beta x^{4.5})\,,
\end{array}\label{eq:UniversalShape}
\end{eqnarray}
where $\mathcal{U}(x)$ is a 'universal' function capturing the essence of the spectral features (peak amplitude and IR/UV slopes), with $\mathcal{U}_1 \equiv \mathcal{U}(1)$ and $\alpha + \beta = 1$. We find $\mathcal{U}_1 \simeq 10^{-5}$ for RD, $\mathcal{U}_1 \simeq 10^{-6}$ for MD, and $\alpha = 0.25, \beta = 0.75$ for both RD and MD. 

The GW energy density spectrum today can be obtained from the spectrum computed at the time of production. Redshifting the amplitude and wavenumbers, we find $h^2\Omega^{(0)}_\GW(f) $ = $h^2\Omega_{\rm rad}^{(0)}(g_0/g_{\rm F})^{1/3}\times({H_{\rm I}/M_p})^{{4}}\times\epsilon_{\rm I}\,q^{1.55}\,\mathcal{U}(k/k_p)$ and $f = \epsilon_{\rm I}^{1/4}\times(k/\rho_{\rm I}^{1/4})\times5\cdot10^{10}~{\rm Hz}$, where $\epsilon_{\rm I} \equiv (a_{\rm I}/a_{\rm RD})^{(1-3w)} \leq 1$, $h^2\Omega^{(0)}_{\rm rad}$ is the fractional energy in radiation today, and $(g_0/g_{\rm F})$ is the ratio of relativistic species today to those at $t_{\rm F}$. Using $\rho_{\rm I} = 3\lambda\varphi_{\rm I}^2M_p^2$, today's frequency $f_p$ and amplitude of the GW background peak $\Omega_\GW^{\rm (p)} \equiv \Omega_\GW(f_p)$ are 
\begin{eqnarray}
\begin{array}{c}
f_p \simeq \epsilon_{\rm I}^{1/4}\, y^{1/2}\left({\varphi_{\rm I}/M_p}\right)^{1/2}\times \,5\cdot 10^{10}\,{\rm Hz}\,,\vspace*{1mm}\\
h^2\Omega_\GW^{\rm (p)} \simeq \epsilon_{\rm I}\,\mathcal{U}_1\,q^{1.55}\left({H_{\rm I}/M_p}\right)^{4}\times 10^{-6}\,,
\end{array}\label{eq:PeakAmplFreq}
\end{eqnarray}
where we used $h^2\Omega_{\rm rad}^{(0)} (g_0/g_{\rm F})^{1/3} \simeq 10^{-6}$. Eqs.~(\ref{eq:PeakAmplFreq}) describe the peak of the GWs from a single fermion species with Yukawa coupling strength $y$. In the SM every charged fermion couples directly to the Higgs, each with a different Yukawa coupling strength, $y_{t} > y_{b} > y_{\tau} > y_{c} > y_{\mu} \gtrsim y_{s} > y_{d} > y_u > y_e$, the labels standing for the quarks $\lbrace t,b,s,c,u,d\rbrace$ and charged leptons $\lbrace e,\mu,\tau\rbrace$. The derivation of Eqs.~(\ref{eq:GW_spectra_Reg}) actually relies on computing an unequal-time-correlator of the type $\sim \langle T_{ij}T_{ij}\rangle$~\cite{Figueroa:2013vif}, assuming that only one fermion species contributes to the energy momentum tensor $T_{ij}$. However, in our case there is a sum over all the fermion species $T_{ij} = \sum_a T_{ij,a}$, so that $\langle T_{ij}T_{ij}\rangle$ = $\sum_{a} \langle T_{ij,a}T_{ij,a}\rangle$ + $\sum_{a\neq b} \langle T_{ij,a}T_{ij,b}\rangle$. Since the creation/annihilation operators of different species anticommute, the cross-terms vanish. This implies that Eqs.~(\ref{eq:GW_spectra_Reg}) and, consequently, Eqs.~(\ref{eq:UniversalShape}),(\ref{eq:PeakAmplFreq}), are valid for each species individually. The total GW spectrum is then a superposition of each individual species' spectra, 
\begin{equation}
h^2\Omega_\GW^{(0)}(f) \simeq \epsilon_{\rm I}\,10^{-6}\,({H_{\rm I}/M_p})^{{4}}\,{\sum}_{a} q_a^{1.55}\,\mathcal{U}(q_a^{-{1/4}}\kappa)\,,
\end{equation}
with $q_a \equiv y_a^2/\lambda_{\rm I}$. Had the amplitude of the peaks scaled as $\Omega_\GW^{(p)} \propto q_a^{r}$ with $r \ll 1$, a series of peaks would emerge in the final spectrum, one peak per fermion. The presence of these peaks would represent a method for probing particle couplings, i.e.~a 'spectroscopy' of particle physics. However, the real scaling of the peaks amplitude as $\propto q_a^{1.55}$, implies that the IR tail of the highest peak completely dominates over the amplitude of the lower peaks, see Figure~\ref{fig:FigureI}. Given the Yukawa coupling strengths of the SM, the amplitudes of each species peak are in proportion $\Omega_\GW^{(p)}\big|_{\rm t}$ $:$ $\Omega_\GW^{(p)}\big|_{\rm b}$ $:$ $\Omega_\GW^{(p)}\big|_{\tau} : ... $ $=$ $q_t^{1.55}$:$q_b^{1.55}$:$q_\tau^{1.55}:...$, located at frequencies $f_p^{(t)}:f_p^{(b)}:f_p^{(\tau)}:...$ = $y_t^{1/2}:y_b^{1/2}:y_\tau^{1/2}:..$. The peak of the top quark dominates the signal overtaking the lower peaks, what makes inaccesible the information on the other species' couplings. 

To compute the frequency $f_p^{(t)}$ and amplitude $h^2\Omega_\GW^{\rm (p)}\big|_{\rm t}$ of the top quark peak today, we need first to fix the resonance parameter $q_t = y_t^2/\lambda_{\rm I}$ at the energy scale $E_{\rm I}$. The Yukawa coupling $y_t$ runs very mildly from $\sim 0.9$ to $\sim 0.4$, between $\sim 10^2$ GeV and $\sim 10^{19}$ GeV, so we can set $y_t(E_{\rm I}) \sim 0.5$ as a representative value. The resonant parameter is then $q_t \sim \mathcal{O}(0.1)/\lambda_{\rm I} \gg 1$, for instance $q_t \sim 10^{6}$ if $\lambda_{\rm I} \sim 10^{-7}$. The smaller $\lambda_{\rm I}$ the bigger $q_t$, and hence the higher the GW peak amplitude. Using the fact that $\varphi_{\rm I} = (a_{\rm e}/a_{\rm I})\varphi_{\rm e} \simeq 0.1(a_{\rm e}/a_{\rm I})H_{\rm e}/\lambda^{1/4}$ and assuming a RD scenario immediately after inflation (i.e.~$\epsilon_{\rm I} = 1$), we find $f_p^{(t)} \sim (H_{\rm e}/H_{\rm e}^{<})^{1/2}\times10^{7}\,{\rm Hz}$, and $h^2\Omega_\GW^{\rm (p)}\big|_{\rm t} \sim \mathcal{U}_1\,10^{-25}(H_{\rm e}/H_{\rm e}^{<})^{4}\,\lambda^{-1.55}$, with $H_{\rm e}^{<} = 9\cdot 10^{13}$ GeV the actual upper bound on the inflationary Hubble scale~\cite{Ade:2013uln}. The lower $H_{\rm e}$ the smaller $f_p^{(t)}$, shifting the GW peak towards the observable low-frequency window of currently planned detectors. However, lowering $H_{\rm e}$ also supresses significantly the amplitude of the signal, which scales as $\propto (H_{\rm e}/H_{\rm e}^{<})^4 \ll 1$. Therefore, $H_{\rm e} \lesssim H_{\rm e}^{<}$ is the only situation at which the peak amplitude might not be strongly supressed. In that case, we still need $\lambda_{\rm I}$ to be very small to reach a sufficiently high peak amplitude. For instance, $\lambda_{\rm I} \lesssim 10^{-7}, 10^{-10}, 10^{-13}$ are needed, to achieve $h^2\Omega_\GW^{\rm (p)}\big|_{\rm t} \gtrsim 10^{-20}, 10^{-15}, 10^{-10}$, respectively. In summary, we see that only if the SM is stable but is extremelly close to the instability region, i.e.~$0 < \lambda_{\rm I} \lll 1$, does the peak signal of the GWs from the quark top have a significant amplitude.\\

\begin{center}
{\bf IV. What if the Higgs was the Inflaton?}\vspace*{-2mm}
\end{center}
In the Higgs-inflation scenario a non-minimal coupling to the Ricci scalar $2\xi R|\Phi|^2$, allows the Higgs to play the role of the inflaton~\cite{Bezrukov:2007ep}. An intense debate is currently ongoing about the viability of this scenario, but we will not enter into this matter here. Instead, we will simply compute the GW production after inflation, assuming the validity of the model. In that case the Higgs oscillates around zero following the end of inflation. In the Einstein frame, writing the Higgs in the unitary gauge $\Phi = \varphi/\sqrt{2}$ and redefining its amplitude as $h = a^{3/2}(3\xi/4)(\varphi/M_p)^2$, it is found~\cite{Bezrukov:2008ut,GarciaBellido:2008ab} that the Higgs oscillates as $h \simeq \sin(\tau)/\tau$, with $d\tau \equiv a(t)Mdt$, $M \equiv M_p/\sqrt{3}r$ the effective mass of the Higgs, and $r \equiv \xi/\lambda^{1/2}$. The Higgs pressure averages to zero over the oscillations, so the universe expands effectively as in MD. 

A background of GWs is generated after the end of inflation, again due to the non-perturbative decay of the Higgs, which corresponds to preheating in this scenario. From the Yukawa interactions, fermions acquire an effective mass in the Einstein frame given by $m_\psi(\tau) = q^{1/2}h^{1/2}(\tau)\,M$, with $q \equiv 2r^2(y^2/\xi)$ a resonance parameter, and $y$ the Yukawa coupling of the given species. Using this effective mass, we can solve the corresponding fermion mode equations, choosing again initial conditions corresponding to vanishing fermion number density. To compute the GW spectrum $\Omega_\GW(k,t;q)$, we simply need to insert the new mode functions $u_{k,\pm}(t)$ into Eq.~(\ref{eq:GW_spectra_Reg}). Following the analysis of Section~III, we find that fermions are excited up to a cut-off scale, this time given by $k_* \sim j^{1/3}q^{1/3}M$, with $j$ the number of Higgs zero-crossings since the end of inflation. Considering that fermion production ends after $j_{\rm F}$ zero-crossings, we find the amplitude and frequency of the GW peak today, for a given fermion species, given by
\begin{eqnarray}
\begin{array}{c}
f_p \simeq \epsilon_{\rm I}^{1/4}\,j_{\rm F}^{1/3}\,q^{1/3}\,r^{-{1/2}}\times \,2\cdot 10^{10}\,{\rm Hz}\,,\vspace*{1mm}\\
h^2\Omega_\GW^{\rm (p)} \simeq \epsilon_{\rm I}\,\mathcal{U}_1\,q^{1.7}\,r^{-4}\times 10^{-7}\,,
\end{array}\label{eq:PeakAmplFreq_HI}
\end{eqnarray}
where $\epsilon_{\rm I} \equiv (a_{\rm I}/a_{\rm RD}) < 1$, whilst the $q^{1.7}$ scaling and amplitude $\mathcal{U}_1 \simeq 10^{2}$ are found from a numerical fit. A 2-loop analysis of the running of the parameters in this model~\cite{Bezrukov:2009db} shows that, for the alowed $125-126$ GeV Higgs mass range, $\xi \sim \mathcal{O}(10^3)$ and $r \sim 5\cdot 10^4$ at the energy scale of inflation. Besides, in~\cite{Bezrukov:2008ut,GarciaBellido:2008ab,Bezrukov:2009db,GarciaBellido:2011de} it has been shown that the Higgs transfers efficiently its energy into the decay products after $\mathcal{O}(100)$ zero-crossings. Finally, note that we can estimate $\epsilon_{\rm I}$ as $\sim j_{\rm RD}^{-2/3}$, with $j_{\rm RD}$ ($\gtrsim j_{\rm F}$) the number of Higgs zero-crossings until RD. Putting everything together, the frequency of each peak today is estimated as $f_p \simeq 2\,y^{2/3}\times10^{10}\,{\rm Hz}$, were we used as fiducial values $\xi = 1000$, $r = 5\cdot 10^4$, $j_{\rm RD} \sim j_{\rm F} = 100$, and $\epsilon_{\rm I}^{1/4}j_{\rm F}^{1/3} \sim j_{\rm RD}^{1/6} \simeq 2$. The GW peaks are in a proportion $f_p^{(a)}:f_p^{(b)} = y_a^{2/3}:y_b^{2/3}$, with $y_a, y_b$ the Yukawa couplings of different species. However, as in the Higgs spectator scenario, the GW peak from the most strongly coupled species -- the quark top -- dominates over the rest of peaks. Therefore, only the peak associated to the top quark remains in the final spectrum of GWs, located at $f_p^{(t)} \sim 10^{10}$ Hz. Choosing the previous fiducial values for $\xi$, $r$ and $\epsilon_{\rm I}$, the amplitude of the peak today, is estimated as $h^2\Omega_\GW^{\rm (p)}\big|_{t} \simeq \mathcal{U}_1\,y_t^{3.4}\times 10^{-15} \sim 10^{-14}$.

\begin{center}
{\bf V. Discussion and Conclusions}\vspace*{-0mm}
\end{center}
A number of aspects not considered in our derivations, might have an impact on the results. The most relevant aspect is the parametric excitation of the $SU(2)_{\rm L}$ gauge vectors $Z,W^{\pm}$, from which new peaks are expected to appear in the GW spectrum. On general grounds, these peaks should be higher than the fermionic ones, since bosons can grow in amplitude arbitrarily, but fermions cannot. However, in the absence of lattice simulations considering the non-linearities and charge currents in the bosonic sector, we will not attempt to estimate their peak amplitude. Let us observe, nonetheless, that given the fact that the $SU(2)_{\rm L}$ gauge coupling is $g_2 \sim y_t$, the GW peaks from the gauge bosons will be located at similar frequencies as that of the top quark, most likely not being possible to resolve them separately. The $SU(2)_{\rm L}$ gauge bosons might therefore enhance the amplitude of the final single peak in the GW spectrum, but we leave the study of this for future research. 

Another relevant aspect is the fermion decay width, which for the top quark is $\Gamma_{\rm t} \sim $ $\mathcal{O}(10^{-3})g_2^2(m_t/m_W)^2m_t$, $m_{\rm t} = y_t\varphi/\sqrt{2}$, $m_W = g_2\varphi/2$. The GWs are created in a step manner only, during the brief periods of fermion non-perturbative excitation $\Delta t \ll T_\varphi$, when the Higgs crosses around zero (twice per oscillating period $T_{\varphi}$). The GW production will not be affected by the top decay unless $\Gamma_{\rm t}\Delta t > 1$. In the Higgs spectator scenario, $\sqrt{\lambda_{\rm I}}\varphi_{\rm I}\Delta t \sim q^{-1/4}$, and the Higgs amplitude during that time is $|\varphi| \leq \varphi_* = q^{-1/4}\varphi_{\rm I}$, so $\Gamma_{\rm t}\Delta t$ $\lesssim$ $\mathcal{O}(10^{-3})$ $\times(y_t^2/q^{1/4})(y_t/\sqrt{\lambda_{\rm I}})(|\varphi|/\varphi_{\rm I})$ $\lesssim$ $\mathcal{O}(10^{-3})y_t^2$ $\ll 1$. The top decay therefore does not affect the GW production. Similar conclusions follow in the Higgs-Inflation case. 

Other aspects that could impact on the final details are the fermions' backreaction onto the Higgs, the possible thermal coupling of the Higgs, and the neglect of quantum corrections in the fermion dynamics. 

Let us also stress the fact that the generation of GWs from non-perturbatively excited fields can also be expected in beyond the SM scenarios. For instance if the Higgs couples to non-SM fields, say to species heavier than the top quark, right-handed neutrinos, etc. Alternatively, we can also concieve an oscillatory scalar field $\phi$ other than the SM Higgs, coupled to either SM or non-SM fields. The single peak in the final GW spectrum will then probe the coupling of the most strongly interacting particle with the oscillatory field. The corresponding GW backgrounds, if detected, would provide a methodology for probing couplings at energies much higher than what any particle accelarator will ever reach.

Summarizing, in this letter we predict that a background of GWs is created due to the non-perturbative decay of the SM Higgs after inflation. The existence of this background and the location of its spectral features should be considered as a robust prediction, though the final details might be affected by the inclusion of the mentioned effects above, to be investigated elsewhere. The GW spectral features could be used for spectroscopy of elementary particles in/beyond the SM, probing at least the coupling of the most strongly interacting species. For this, new high frequency GW detection technology must be developed, beyond that currently planned~\cite{Cruise:2006zt,Akutsu:2008qv,Cruise:2012zz}.
 

{\it Acknowledgements}. I am very grateful to Julian Adamek, Diego Blas, Ruth Durrer, Jorge Nore\~na, Subodh Patil, and Toni Riotto, for providing useful comments and criticism on the manuscript, and to Juan Garc\'ia-Bellido, Kimmo Kainulainen, Jos\'e M. No, Marco Peloso, Javier Rubio, Sergey Sibiryakov and Lorenzo Sorbo, for useful discussions on different technical aspects of the subject. Special gratitude goes to Kari Enqvist and Tuukka Meriniemi, with whom I first studied the GW production from fermions. I would also like to thank Matteo Biagetti, Kwan Chuen Chan, Enrico Morgante, Azadeh Moradinezhad-Dizgah, Ermis Mitsou, and all the Basel, CERN, UniGe, Nikhef and IFT crowds. Last, but not least important, I am also in debt with Jimmy Page, Robert Plant, Malcom Young and Angus Young, for suggesting me very nice titles for the paper, and providing me with a great inspiration. This work has been supported by the Swiss National Science Foundation.


\bibliography{Stairway2Heaven}

\end{document}